\documentclass[aps,prb,reprint,superscriptaddress]{revtex4-2}

\usepackage{graphicx, amsmath, amssymb, mhchem, xspace, braket}
\usepackage{color}
\usepackage{lineno,hyperref}

\newcommand{\sub}[1]{\ensuremath{_{\textrm{#1}}}}
\newcommand{\SSO}{\ce{Sr3SnO}\xspace}
\newcommand{\SxSO}{\ce{Sr_{3-x}SnO}\xspace}
\newcommand{\StwoSO}{\ce{Sr2Sn2O2}\xspace}
\newcommand{\SthreeSO}{\ce{Sr3Sn2O2}\xspace}
\newcommand{\SfourSO}{\ce{Sr4Sn2O2}\xspace}
\newcommand{\Sn}{\ce{^{119}Sn}\xspace}
\newcommand{\APO}{\ce{\textit{A}3\textit{B}O}\xspace}
\newcommand{\PO}{\ce{\textit{AB}O3}\xspace}
\newcommand{\rev}[1]{#1}
\newcommand{\revrev}[1]{#1}

\begin{document}


\title{Negative ionic states of tin in the oxide superconductor\\\SxSO revealed by M\"{o}ssbauer spectroscopy}


\author{Atsutoshi Ikeda}
\email[]{a.ikeda@scphys.kyoto-u.ac.jp}

\author{Shun Koibuchi}
\affiliation{Department of Physics, Kyoto University, Kyoto 606-8502, Japan}

\author{Shinji Kitao}
\affiliation{Institute for Integrated Radiation and Nuclear Science, Kyoto University, Osaka 590-0494, Japan}

\author{\\Mohamed Oudah}
\affiliation{Department of Physics, Kyoto University, Kyoto 606-8502, Japan}
\affiliation{Stewart Blusson Quantum Matter Institute, University of British Columbia, Vancouver BC, V6T 1Z4 Canada}

\author{Shingo Yonezawa}
\affiliation{Department of Physics, Kyoto University, Kyoto 606-8502, Japan}

\author{Makoto Seto}
\affiliation{Institute for Integrated Radiation and Nuclear Science, Kyoto University, Osaka 590-0494, Japan}

\author{Yoshiteru Maeno}
\affiliation{Department of Physics, Kyoto University, Kyoto 606-8502, Japan}


\date{\today}

\begin{abstract}
We report the temperature variation of the \Sn-M\"{o}ssbauer spectra of the antiperovskite (inverse perovskite) oxide superconductor \SxSO.
Both superconductive (Sr-deficient) and non-superconductive (nearly stoichiometric) samples exhibit major $\gamma$-ray absorption
with isomer shift similar to that of \ce{Mg2Sn}.
This fact shows that \SxSO contains the metallic anion \ce{Sn^{4-}}, which is rare especially \revrev{among} oxides.
In both samples, we observed another $\gamma$-ray absorption with a larger isomer shift, 
indicating that there \revrev{is another ionic state} of Sn with \revrev{a} higher oxidation \revrev{number}.
The temperature dependence of the absorption intensities reveals that the Sn ions exhibiting larger isomer shifts have a lower energy of the local vibration.
The larger isomer shift and lower vibration energy are consistent with the values estimated from
the first-principles calculations for hypothetical structures with various Sr-deficiency arrangements.
Therefore, we conclude that the additional $\gamma$-ray absorptions originate from the Sn atoms neighboring the Sr deficiency.
\end{abstract}


\maketitle

\section{Introduction\label{introduction}}
\begin{figure}
\includegraphics[width=0.75\linewidth]{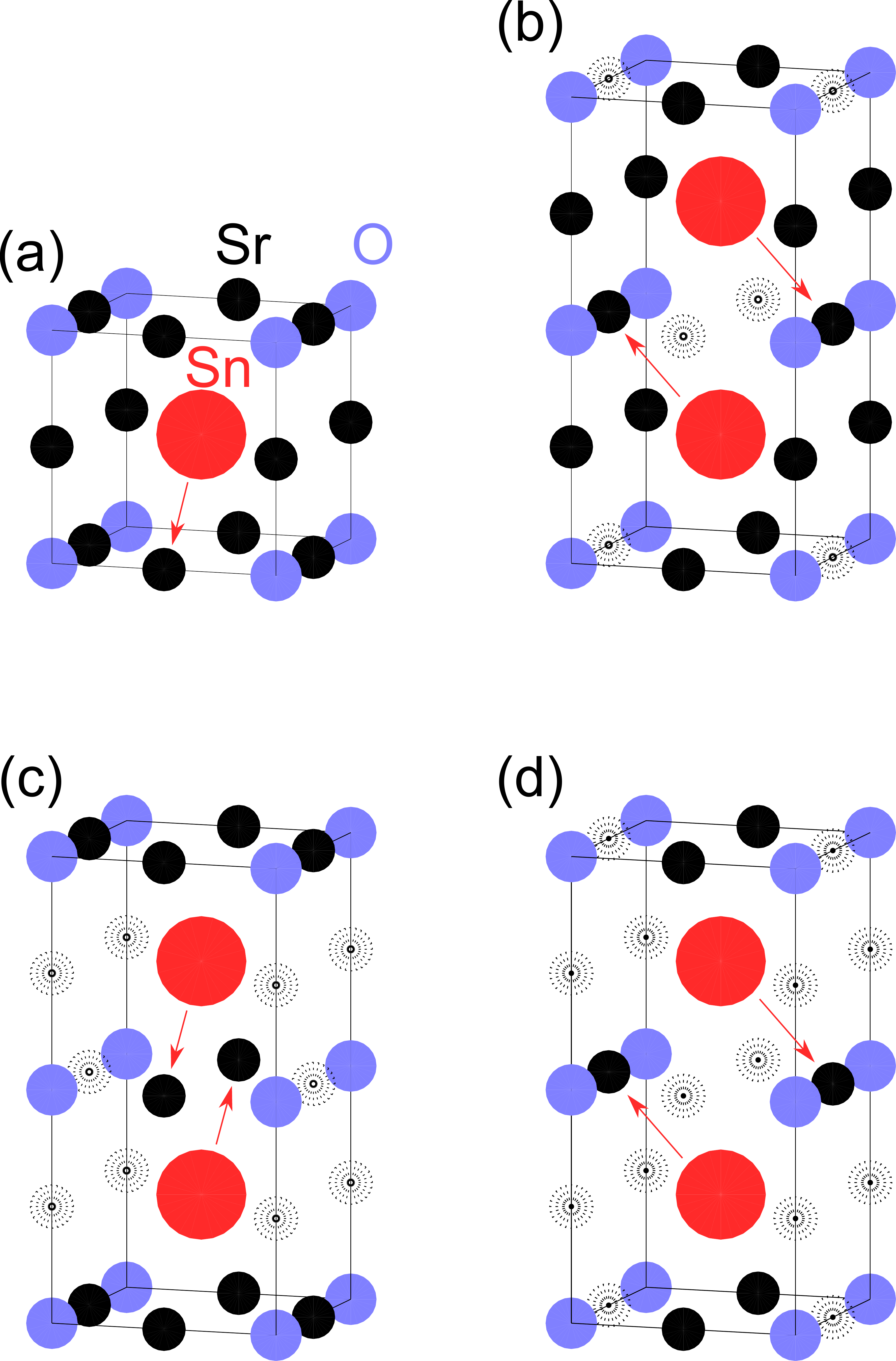}
\caption{(Color online) (a) Crystal structure of \SSO.
In contrast to the common picture with the \ce{OSr6} octahedron \revrev{shown} at the center, 
Sn (red sphere) is placed at the center of the unit cell in this picture.
(b)--(d) Three hypothetical structures of highly Sr-deficient antiperovskite oxides (b) ``\SfourSO,'' (c) ``\SthreeSO,'' and (d) ``\StwoSO.''
The dotted spheres indicate the assumed positions of the Sr deficiency.
The arrows indicate the direction of the atomic shift for calculation of the force constants of Sn vibrations.
The figures were prepared with the program VESTA~\cite{Momma2011VESTA}.}
\label{fig: crystalline_structures}
\end{figure}

Antiperovskite (inverse perovskite) oxides \APO (\textit{A}: alkaline-earth elements, Eu, or Yb; \textit{B}: group 14 elements) 
are the metal-rich counterpart of the ordinary perovskite oxides \PO, with inverted positions of the positive and negative ions~\cite{widera1980ubergangsformen}.
In antiperovskite oxides, an oxygen ion is octahedrally coordinated by the metallic \ce{\textit{A}^{2+}} ions [Fig.~\ref{fig: crystalline_structures}(a)]; 
while in ordinary perovskite oxides, a metal element is surrounded by \ce{O^{2-}}.
As a result of three \ce{\textit{A}^{2+}} ions in a unit cell, the oxidation number of \textit{B} is forced to be $4-$ such as \ce{Sn^{4-}} and \ce{Pb^{4-}} 
to satisfy the charge neutrality as \ce{$\left(A^{2+}\right)$3\textit{B}^{4-}O^{2-}}.
Novel physics and chemistry of such metallic anions, especially rare in oxides, \revrev{have been motivating} us to study this group of materials.
Since M\"{o}ssbauer spectroscopy is applicable to the Sn nucleus, microscopic characterization of such a \ce{Sn^{4-}} state 
may unveil the nature of electronic states of antiperovskite oxides.

\revrev{Due to the recent predictions of b}ulk Dirac cones~\rev{\cite{kariyado2011three, kariyado2012low}} 
and topological nature~\rev{\cite{hsieh2014topological}} in the vicinity of the Fermi energy\revrev{,} 
antiperovskite oxides are attracting a lot of attention~\cite{kariyado2011three, kariyado2012low, hsieh2014topological, Nuss2015tilting, 
Okamoto2016thermoelectric, oudah2016superconductivity, Obata2017Ca3PbOARPES, Kariyado2017Ba3SnO, Ikeda2018bandstructure, 
Niklas2018Sr3-xSnO, Suetsugu2018Sr3PbO, Kawakami2018TopoSC, Kitagawa2018Sr3-xSnO, Oudah2019SrDeficiency, Kariyado2017pseudoLandauLevels}.
%
After the initial theoretical predictions, the chemical and physical characters of antiperovskite oxides have been extensively investigated~\cite{Nuss2015tilting}.
The metallic anion \ce{Sn^{4-}} was experimentally observed in \ce{Sr3SnO} using the M\"{o}ssbauer spectroscopy 
at room temperature by some of the present authors~\cite{Oudah2019SrDeficiency}.
The Dirac cone in the bulk band structure is confirmed in \ce{Ca3PbO} with the angle-resolved photoemission spectroscopy (ARPES)~\cite{Obata2017Ca3PbOARPES}, 
in \ce{Sr3PbO} with magnetotransport~\cite{Suetsugu2018Sr3PbO}, and in \ce{Sr3SnO} with nuclear magnetic resonance (NMR)~\cite{Kitagawa2018Sr3-xSnO}.
Furthermore, thermoelectric properties of \ce{Ca3SnO} and \ce{Ca3Pb_{1-x}Bi_xO} are studied 
to make use of the multivalley band structure with six equivalent Dirac cones~\cite{Okamoto2016thermoelectric}.
\rev{\revrev{More recently, it has been proposed that} one can tune the size of the band inversion and 
mass of the Dirac cone via chemical substitution of \textit{A} and \textit{B}~\cite{Kariyado2017Ba3SnO}.
Thus, antiperovskite oxides are a good platform to study the Dirac and topological natures.}

In 2016, some of the present authors discovered superconductivity in \SxSO 
prepared with the nominal Sr deficiency of $x_0 = 0.5$~\cite{oudah2016superconductivity}, 
the first superconductivity among the antiperovskite oxides.
This material exhibits two superconducting transitions at $T\sub{c} \simeq 5$~K and $0.8$~K\@.
Theoretical analyses indicate that the superconductivity in \SxSO \revrev{can be} topological crystalline superconductivity with $j = 3/2$ pairing 
reflecting the topology of the electronic band structure in the normal state~\cite{oudah2016superconductivity, Kawakami2018TopoSC}.
Band structure calculations~\cite{Ikeda2018bandstructure} and NMR experiments~\cite{Kitagawa2018Sr3-xSnO} 
suggest heavy hole doping to Sn \revrev{due to} Sr deficiency.
Indeed, M\"{o}ssbauer spectra at room temperature \revrev{revealed additional} ionic state of Sn \revrev{having less electrons than} \ce{Sn^{4-}}, 
and its fraction increases with Sr deficiency~\cite{Oudah2019SrDeficiency}.
However, the microscopic relation between the Sn states and Sr deficiency was not understood.
In addition, temperature evolution of the M\"{o}ssbauer spectra has not been reported.

In this paper, we present the temperature dependence of the Sn-M\"{o}ssbauer spectra for the nearly stoichiometric and Sr-deficient \SxSO samples.
Both samples exhibit two Sn states; \ce{Sn^{4-}} characteristic of antiperovskite oxides 
and the other with a higher isomer shift particularly visible in the deficient sample.
Detailed analyses of the temperature dependent spectra show 
that the additional Sn state has a smaller number of electrons and lower energy of the local lattice vibrations.
We also performed the first-principles calculations on various hypothetical Sr-deficient \revrev{arrangements} 
and successfully reproduced the experimental results.
Therefore, we conclude that the additional Sn state originates from Sn \revrev{atoms} neighboring the Sr deficiency.


%



\section{Experiment\label{experiment}}
\subsection{Sample Preparation}
Polycrystalline samples of \SxSO were synthesized using the reaction:
\ce{(3$-x_0$)Sr + SnO -> Sr_{3-x}SnO + ($x-x_0$)Sr ^},
where $x_0$ is the nominal Sr deficiency.
Since \ce{Sr} and \SxSO are highly air-sensitive, reactants and products were handled in a glovebox filled with argon.
\ce{Sr} (Sigma-Aldrich, 99.99\%) and \ce{SnO} (Sigma-Aldrich, 99.99\%) 
were put in iron crucibles (Chiyoda Kogyo Seisakusho) with the molar ratio of $(3-x_0)$:1.
The nearly stoichiometric (NS) sample was synthesized with the Sr-excess condition $x_0=-0.5$ 
because preparation with $x_0 = 0$ results in SrO impurity contained in the product.
The crucible was sealed in a stainless-steel capsule 
(Fujikin, SUS316L pipe, $\phi$19.05$\times$1.65t$\times$110~mm sealed with Fujikin, V-LOK Cap, VUWJC-19.05) under 1-atm argon at room temperature.
The sealed capsule was heated to 1100$^\circ$C in 10.5 hours, kept there for 50 hours, and cooled down to 800$^\circ$C in 6 hours, 
using a muffle furnace (Denken Co., Ltd., KDF-S80).
After subsequently kept for 20 hours at 800$^\circ$C, the capsule was cooled to room temperature in 8 hours~\cite{Nuss2015tilting}.
For the Sr-deficient (D) sample prepared with $x_0=+0.5$, the crucible was sealed in a quartz tube under 0.3~atm of argon at room temperature.
The tube was heated up to \revrev{850}$^\circ$C over 3 hours, kept for 3 hours, and quenched in water.
\revrev{Then the tube was heated again at 600$^\circ$C for 48 hours and the heater of the furnace was switched off.}

For both NS \rev{($x\simeq0$)} and D \rev{($x\simeq0.5$)} samples, the actual Sr deficiency $x$ was not measured.
Since the Sr evaporation during the heating process was \revrev{about 0.2\% based on the difference of the mass before and after the reaction}, 
we expect that the actual amount of deficiency is in the D sample close to its nominal value 0.5.
For the NS sample, the evaporation was \revrev{only 6\%}, 
but the excess Sr seems to stick on the wall of the crucible, leading to nearly stoichiometric value ($x\simeq0$).
\rev{For the samples prepared with similar conditions to the NS sample, the energy dispersive x-ray spectroscopy (EDX) indicates 
that the Sr/Sn ratio is around 2.98 (\revrev{i.e.} $x=0.02$).
\revrev{This result supports that the actual $x$ value of the NS sample is indeed small, but also 
suggests that there is} a \revrev{certain} amount of the spontaneous Sr deficiency even in the NS sample.
We note that Ca deficiency and corresponding hole doping are observed in a single crystal of \ce{Ca3PbO} 
by the electron-probe microanalysis (EPMA) and ARPES~\cite{Obata2017Ca3PbOARPES}.
For the samples prepared with the \revrev{similar} conditions to the D sample, EDX shows a substantial spatial distribution of the Sr/Sn ratio, 
with the largest population at Sr/Sn $=2.5$ (\revrev{i.e.} $x=0.5$)~\cite{Niklas2018Sr3-xSnO}.}

\subsection{Characterization}
Powder X-ray diffraction (pXRD) patterns were collected with the \ce{Cu}-$K\alpha$ radiation 
(wavelengths of 0.1540538~nm for $K\alpha_1$ and 0.1544324~nm for $K\alpha_2$) using a commercial diffractometer (Bruker AXS, D8 Advance) 
equipped with an array of 192 detectors.
The diffracted X ray was integrated over 0.3~s for each angle, and the total measurement time was approximately 30 minutes for each sample.
In order to prevent decomposition of the samples during the measurements, 
the samples were placed on a glass plate under argon atmosphere and were covered with a 12.5-$\mu$m-thick polyimide film (Du Pont-Toray CO., LTD, Kapton 50H) 
fixed with vacuum grease (Dow Corning Toray, High-Vacuum Grease).

\subsection{Magnetic Properties}
The size of the magnetic moment was measured using a commercial magnetometer with a superconducting quantum interference device (Quantum Design, MPMS-XL).
Powder sample was put in a \revrev{thin} plastic capsule under argon atmosphere to avoid exposure to air.

\subsection{M\"{o}ssbauer spectroscopy}
\ce{^{119\textrm{m}}Sn} in the form of \ce{CaSnO3} (Ritverc GmbH, 740~MBq) was used as the $\gamma$-ray source.
The velocity of the source was calibrated using the absorptions of \ce{^{57}Fe} and its origin was set to the isomer shift of \ce{BaSnO3} at room temperature.
A palladium film with the thickness of 75~$\mu$m was placed between the source and sample in order to shut the X-ray fluorescence of tin.
After the measurements \revrev{of magnetization}, the powder (around 40~mg) was taken out from the capsule and 
mixed with boron nitride (Kishida Chemical, 99.5\%; approximately 70~mg) 
and polyethylene (Beckman RIIC, polyethylene powder for IR spectroscopy; approximately 8~mg) powders under nitrogen atmosphere in order to improve the spatial homogeneity of the sample.
Then the sample was pressed into a pellet with a diameter of 10~mm 
and sealed in multiple layers of polyethylene bags with the thickness of 0.1~mm in order to avoid direct contact to air.
All the measurements, from room temperature to below 3~K, were carried out inside a \ce{^4He} cryostat (Janis Research, ST-400).
Typical measurement time was 12 hours at each temperature.

\subsection{First-Principles Calculation}
\revrev{To analyze} the M\"{o}ssbauer spectra of \SxSO, we performed first-principles calculations.
\revrev{Here, we calculated electronic densities of various \revrev{Sn-based} materials including \SxSO 
in order to deduce the relation between the electronic density and the isomer shift.}
The density of electrons at the tin nucleus position was calculated 
by the full-potential linearized augmented plane-wave plus local orbitals method using the WIEN2k package~\cite{blaha2016wien2k}.
The Perdew-Burke-Ernzerhof generalized gradient approximation~\cite{Perdew1996PBE} was selected as the exchange-correlation functional.
The spin-orbit coupling was taken into account.
We chose the \revrev{muffin}-tin radius (RMT) of each atom to be RMT$\sub{Sr}=2.41$, RMT$\sub{Sn}=2.5$, and RMT$\sub{O}=2.41$ in the unit of the Bohr radius $a_0$.
We set the plane-wave cut off as $RK\sub{max}=8$, the highest angular momentum as $l\sub{max}=10$, 
maximum magnitude of the largest vector in charge density Fourier expansion as $G\sub{max}=18$, 
and the separation energy between the valence and core states as $-7.0$~Ry~\cite{Batool2017Sr3SnOCalculation}.
Only when estimating the energy of the local lattice vibrations in the hypothetical superstructure ``\ce{Sr2Sn2O2},'' 
we used the separation energy of $-6.0$~Ry in order to avoid a technical problem.
The Inorganic Crystal Structure Database (ICSD) numbers of the experimentally reported structures and the sizes of the $k$ mesh used for calculations were
78894 and $12\times12\times12$ for \ce{SnF4}, 
239582 and $12\times12\times12$ for \ce{BaSnO3}, 
411242 and $10\times15\times10$ for \ce{SnCl4}, 
43594 and $14\times14\times7$ for \ce{SnSe2}, 
642850 and $12\times12\times12$ for \ce{Mg2Sn}, 
40038 and $12\times12\times12$ for $\beta$-\ce{Sn}, 
186650 and $6\times17\times16$ for \ce{SnSe}, 
15452 and $10\times18\times8$ for \ce{SnCl2}.
The calculated electronic densities of these compounds were compared 
with the experimentally observed isomer shifts~\cite{Stevens1983Mg2SnMoessbauer, Fournes1986SnF4Moessbauer}.
For \ce{Sr3SnO}, we used the structure reported by J. Nuss \textit{et al.}~\cite{Nuss2015tilting} and the $k$ mesh of $12\times12\times12$.
For strontium-deficient \SxSO, hypothetical structures based on \SSO (``\SfourSO,'' ``\SthreeSO,'' and ``\StwoSO,'' see Fig.~\ref{fig: crystalline_structures}; 
$k$ mesh of $12\times12\times6$) were assumed.

\section{Results\label{results}}
\begin{figure*}
\includegraphics[width=\linewidth]{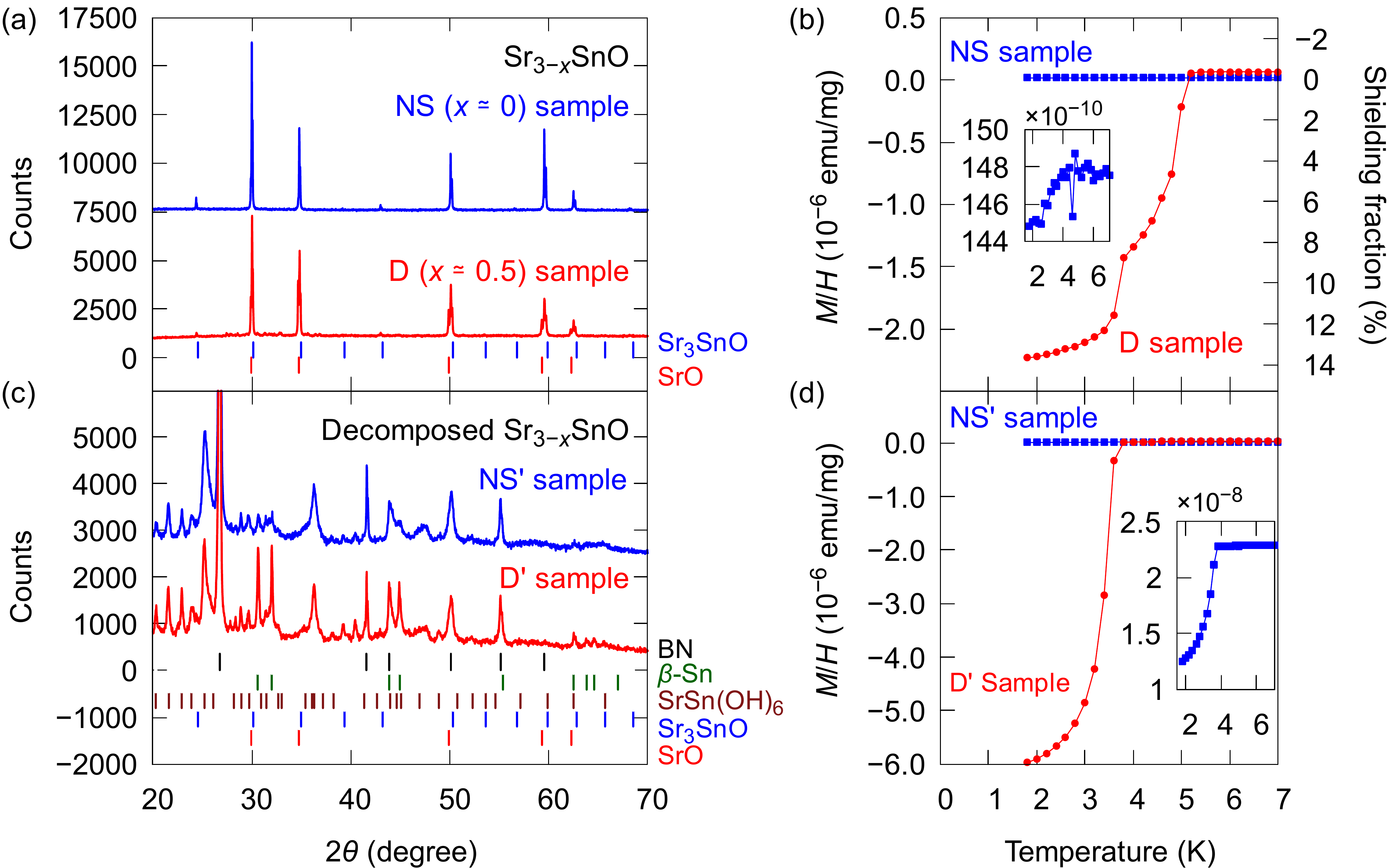}
\caption{(color online) (a) \revrev{Powder x}-ray diffraction patterns 
of the nearly stoichiometric (NS) sample (prepared with the Sr-excess condition of $x_0=-0.5$) 
and Sr-deficient (D) sample (prepared with the Sr-deficient condition of $x_0=+0.5$).
The bars at the bottom indicate the peak positions of \SSO (PDF 01-084-0254) and SrO (PDF 01-075-6979).
(b) Temperature dependence of the magnetization of \SxSO measured in a magnetic field of $\mu_0H=1$~mT under a zero-field-cooling (ZFC) condition.
The inset show a magnified view for the NS sample.
(c) X-ray diffraction patterns after the NS and D samples were intensionally decomposed in air (NS' and D' samples, \revrev{respectively}).
The bars at the bottom indicate the peak positions of BN (PDF 00-034-0421)\revrev{,} \revrev{$\beta$-}Sn (PDF 01-086-2265)\revrev{, \ce{SrSn(OH)6} (PDF 00-009-0086), 
\SSO, and SrO}.
(d) Temperature dependence of the magnetization of the NS' and D' samples measured in a magnetic field of $\mu_0H=2$~mT \revrev{and} under a ZFC condition.
The inset shows a magnified view for the NS' sample. \revrev{When evaluation magnetization, the mass of the mixed BN and polyethylene powder has been subtracted.}}
\label{fig: XRD}
\end{figure*}

\subsection{Sample Characterization}
First, we show pXRD patterns of our samples in Fig.~\ref{fig: XRD}(a).
The NS \rev{($x\simeq0$)} sample exhibits no \revrev{detectable} impurity peaks.
All the peaks were indexed with the space group $Pm\bar{3}m$ (No.\ 221, $O_h^1$) 
and the lattice parameter of $a=0.513767(6)$~nm \revrev{of \SSO~\cite{Nuss2015tilting}}.
For the pattern of the D \rev{($x\simeq0.5$)} sample, the peaks were indexed with $a=0.51246(5)$~nm 
except for the left shoulder peaks originating from insulating \ce{SrO}, 
which also has a cubic crystalline structure (space group No.\ 225, $Fm\bar{3}m$, $O_h^5$) with a slightly larger lattice parameter $a=0.51463(5)$~nm.
Figure~\ref{fig: XRD}(b) shows the temperature dependence of the magnetizations \revrev{of these samples}.
The NS sample does not show superconductivity \revrev{of} \SxSO.
As shown in the inset, the magnetization very slightly decreases below 3.7~K, which originates from the superconductivity 
of tin impurity that was not detected with pXRD\@.
In contrast, the D sample shows the Meissner effect below 5~K, 
originating from the superconductivity of \SxSO~\revrev{\cite{oudah2016superconductivity}}.
The shoulder-like structure at 3.7~K is attributable to the superconductivity of Sn impurity.
\revrev{Comparing magnetization data above and below 3.7~K, we estimate the diamagnetism due to superconductivity of $\beta$-Sn 
as $0.58\times10^{-6}$~emu/mg, which corresponds to 5.4 weight percent or 14 mole percent of $\beta$-Sn inclusion.}

After the magnetization and M\"{o}ssbauer measurements, \revrev{both NS ($x\simeq0$) and D ($x\simeq0.5$)} 
samples were intentionally decomposed in air for a few days 
\rev{[}named hereafter NS' and D' samples\rev{]} and their pXRD patterns and magnetizations were measured again.
The large peak at $2\theta = 27^\circ$ as well as several others in the pXRD pattern [Fig.~\ref{fig: XRD}(c)] 
originate from boron nitride mixed for the M\"{o}ssbauer experiments.
\revrev{The decomposed samples mainly consist of \ce{SrSn(OH)6.}
In addition, t}he D' sample exhibits \revrev{clear} peaks from \revrev{$\beta$-}Sn 
\revrev{whereas the peaks from $\beta$-Sn in} the NS' sample \revrev{are rather faint}.
As shown in Fig.~\ref{fig: XRD}(d), the D' sample exhibits strong diamagnetism below 3.7~K due to the superconductivity of Sn.
The NS' sample also exhibits \revrev{much weaker} diamagnetism as shown in the inset.

\begin{figure*}
\includegraphics[width=\linewidth]{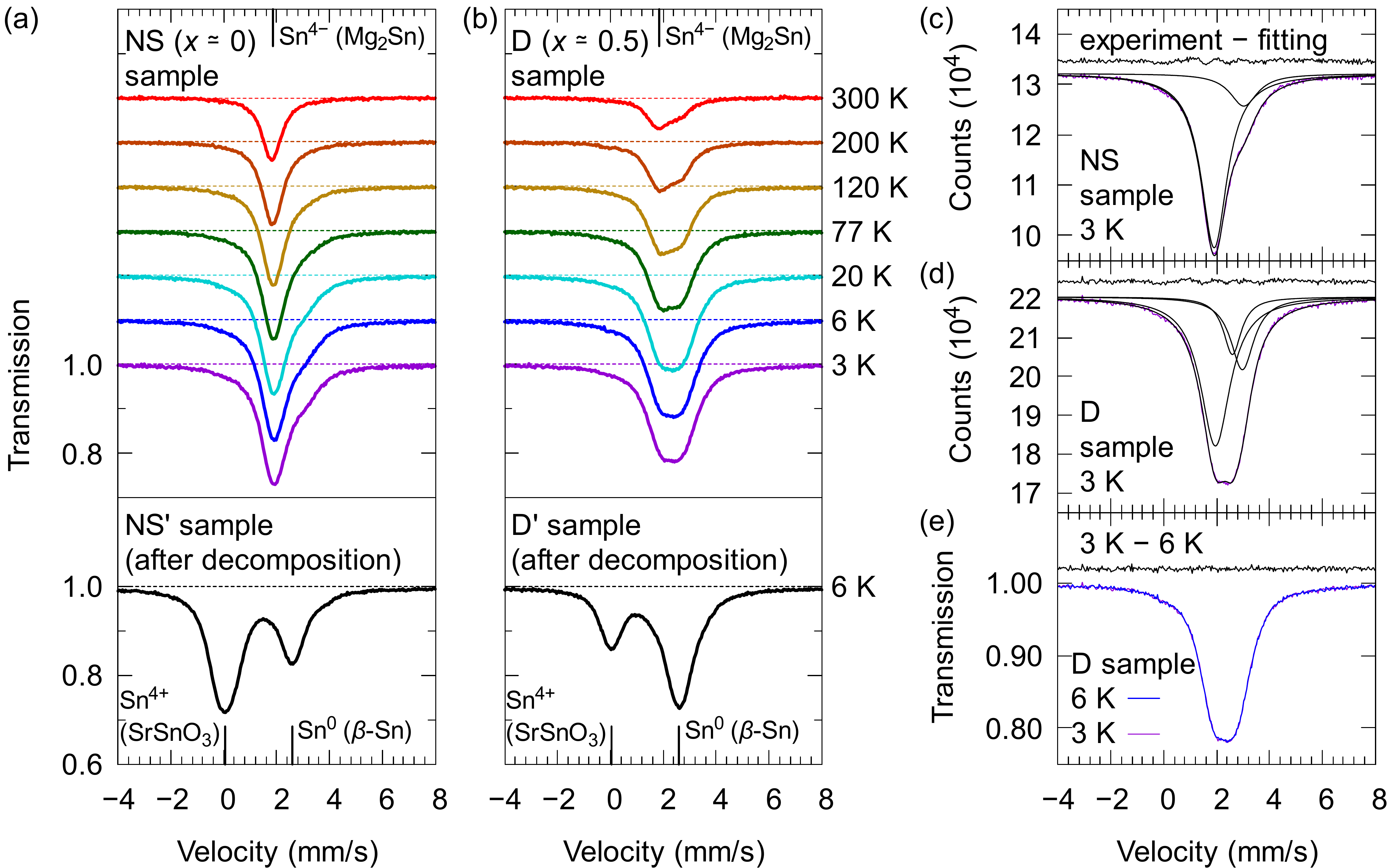}
\caption{(color online) Temperature dependence of the \Sn-M\"{o}ssbauer spectra of (a) the nearly stoichiometric (NS) and (b) Sr-deficient (D) \SxSO samples.
All spectra have the same scale, and each spectrum is 
shifted upward by 0.1 so that the change of the peak shape is more visible.
The dashed line for each spectrum indicates the transmission of 100\%.
For each panel, the bottom spectrum was obtained at 6~K after intensional decomposition in air (NS' and D' samples).
The vertical lines at 1.87~mm/s, 0.05~mm/s, and 2.60~mm/s indicate the isomer shifts of \ce{Mg2Sn} at 300~K~\cite{Stevens1983Mg2SnMoessbauer}, 
\ce{SrSnO3} at 100~K~\cite{Hien1963BaSnO3isomerShift}, and $\beta$-Sn at 4.2~K~\cite{Micklitz1972SnMoessbauer}, 
respectively\revrev{, representing \ce{Sn^{4-}}, \ce{Sn^{4+}}, and \ce{Sn^0} ionic states}.
(c\revrev{, d}) \revrev{Results of} Lorentzian fitting\revrev{s} of the spectra of the NS and D samples 
at 3~K for quantitative analysis of the isomer shifts and absorption intensities.
The sum of the Lorentzians reasonably reproduces the experimental spectr\revrev{a}.
(\revrev{e}) Comparison of the spectra above (6~K, blue curve) and below (3~K, purple curve) the superconducting transition temperature of the D sample.
The two \revrev{spectra almost overlap} each other, and the difference 
\revrev{between the two spectra} (shifted upward) is a straight line within the noise level.}
\label{fig: Moessbauer}
\end{figure*}

\subsection{M\"{o}ssbauer spectroscopy}
\Sn-M\"{o}ssbauer spectra \revrev{of \SxSO samples} at various temperatures are presented in Fig.~\ref{fig: Moessbauer}.
Both \revrev{NS ($x\simeq0$) and D ($x\simeq0.5$)} samples strongly absorb the $\gamma$ ray around the isomer shift of +1.9~mm/s.
This shift is \revrev{very close} to that of \ce{Mg2Sn} (+1.87~mm/s at 300~K~\cite{Stevens1983Mg2SnMoessbauer}), in which the \ce{Sn^{4-}} state is anticipated.
In addition, minor $\gamma$-ray absorptions are observed at slightly higher isomer shifts, as evidenced by the shoulder-like structure in the spectra.

To analyze the data more quantitatively, we fitted the spectra with \revrev{multiple} Lorentzian functions 
and evaluated their isomer shifts and integrated peak intensities.
\revrev{For the NS sample, we used two Lorentzians to fit the major and minor absorptions [Fig.~\ref{fig: Moessbauer}(c)].
We confirmed that the minor absorption is not due to $\beta$-Sn impurity because the observed isomer shift of +3.03(2)~mm/s 
at the lowest temperature is clearly different from the reported isomer shift of $\beta$-Sn (2.60~mm/s at 4.2 K~\cite{Micklitz1972SnMoessbauer}).
For the D sample, we added one extra Lorentzian to take into account the contribution of the $\beta$-Sn impurity [Fig.~\ref{fig: Moessbauer}(d)].
The isomer shift of the Sn impurity is fixed to the reported value.
The intensity is also fixed so that the integrated peak area becomes 14\% 
of the total area at the lowest temperature as calculated from the temperature dependence of the magnetization [Fig.~\ref{fig: XRD}(b)].}
The fitting \revrev{well explains the data}, and the isomer shift of the \revrev{minor} absorption at 3~K 
was evaluated to be \revrev{+2.973(11)}~mm/s\revrev{, which is again distinct from the isomer shift of $\beta$-Sn.
Thus, this minor absorption of the D sample should be intrinsic to \SxSO.}

From the integrated \revrev{intensity} of the spectra at the lowest temperature, the fractions of the tin atoms related to these minor absorptions 
are estimated to be \revrev{16.4(9)}\% in the NS \rev{($x\simeq0$)} sample and \revrev{27.9(12)}\% in the D \rev{($x\simeq0.5$)} sample.
Since the intensity of the minor absorption increases with the strontium deficiency, 
the \revrev{tin} atoms near the Sr deficiency are ascribable to the origin of this minor absorption.

\rev{In order to estimate the amount of the Sr deficiency $x$, let us assume for simplicity 
that the $Z=12$ Sr atoms surrounding a Sn atom can be extracted independently.
In this case, the probability \revrev{that} a Sn atom \revrev{does not have any} neighboring Sr deficiency is $(1-x/3)^Z$.
Then, the relative intensity of the minor absorption, originating from Sn atoms \revrev{next to} Sr deficiencies, is given by $1-(1-x/3)^Z$.
Assuming that the observed \revrev{16}\% minor absorption in the NS $(x\simeq0)$ sample is explained by this formula, we obtain $x=\revrev{0.04}$.
This estimated value of $x$ is consistent with the results of EDX, yielding $x=0.02$ as explained above.
\revrev{This} good agreement confirms \revrev{the scenario} 
that \revrev{minute} spontaneous Sr deficiency lead\revrev{s} to a minor absorption with \revrev{relatively} strong intensity.
Similar amount of the Ca deficiency is reported by EPMA in a single crystal of \ce{Ca3PbO}~\cite{Obata2017Ca3PbOARPES}.
\revrev{Such} amount of deficiency should be insufficient \revrev{to trigger} superconductivity.
For the D ($x\simeq0.5$) sample, \revrev{the same} assumption leads to the equation $1-(1-x/3)^Z=\revrev{0.28}$, whose solution is $x\simeq\revrev{0.08}$.
This value is \revrev{substantially} smaller than the nominal Sr deficiency of $x_0=0.5$, 
probably because the simple assumption of independent deficiency \revrev{distribution} is no longer valid for such a large value of $x$.
This implies that the Sr deficiencies tend to cluster each other, rather than distributing just randomly\revrev{,
when the amount of deficiency becomes larger}}.

Let us discuss the origin of the minor absorption in more detail.
\revrev{Generally speaking, t}he isomer shift reflects the density of electrons at the nucleus position.
\rev{This is because the nuclei in the ground and excited states have slightly different effective radii.
For the Sn nucleus, the excited nucleus has a larger radius than the ground-state one.
Thus, if the density of the electrons at the nucleus site is larger, the energy difference between the \revrev{excited} and ground states becomes larger
because the excited nucleus feels additional Coulomb \revrev{repulsion} by the surrounding electrons.
As a consequence, a larger density of electrons results in a larger isomer shift}~\cite{Fujita1999Moessbauer}.
Thus, more $s$ electrons lead to a more positive isomer shift because $s$ electrons have non-zero probability amplitude at the nucleus position.
In contrast, $p$ or $d$ electrons result in a negative and smaller shift via \revrev{the} screening effect.
Therefore, \revrev{it is expected that} the minor absorption with \revrev{a} larger isomer shift 
originates from tin ions with less $p$ electrons than \ce{Sn^{4-}};
or in other words, the hole-doped ionic state\rev{s} due to strontium deficiency.

In order to confirm that the minor absorptions do not originate from partial decomposition of the samples, 
we measured M\"{o}ssbauer spectra of the intentionally decomposed samples \rev{[}NS' and D' samples\rev{]}.
After the decomposition, the $\gamma$-ray absorptions are located at 0.0--0.1~mm/s and 2.6~mm/s at 6~K\@.
These absorptions are very similar to those of \ce{SrSnO3} (0.05~mm/s at 100~K~\cite{Hien1963BaSnO3isomerShift}) 
and $\beta$-Sn (2.60~mm/s at 4.2~K~\cite{Micklitz1972SnMoessbauer}).
This result indicates that \SxSO decomposes into \revrev{$\beta$-Sn and a certain material containing \ce{Sn^{4+}} ion 
[presumably \ce{SrSn(OH)6} as seen in the pXRD pattern]}, and possibly some other compounds without tin.
\rev{If the samples had decomposed during the setup of the M\"{o}ssbauer experiments, 
we would \revrev{have} observe\revrev{d} the absorption of \revrev{\ce{Sn^{4+}}}.
However, the actual spectra \revrev{of the pristine samples} do not show any absorption at around 0~mm/s.
Therefore, we can exclude the possibility that the minor absorptions originate from partial decomposition.}

\begin{figure}
\includegraphics[width=\linewidth]{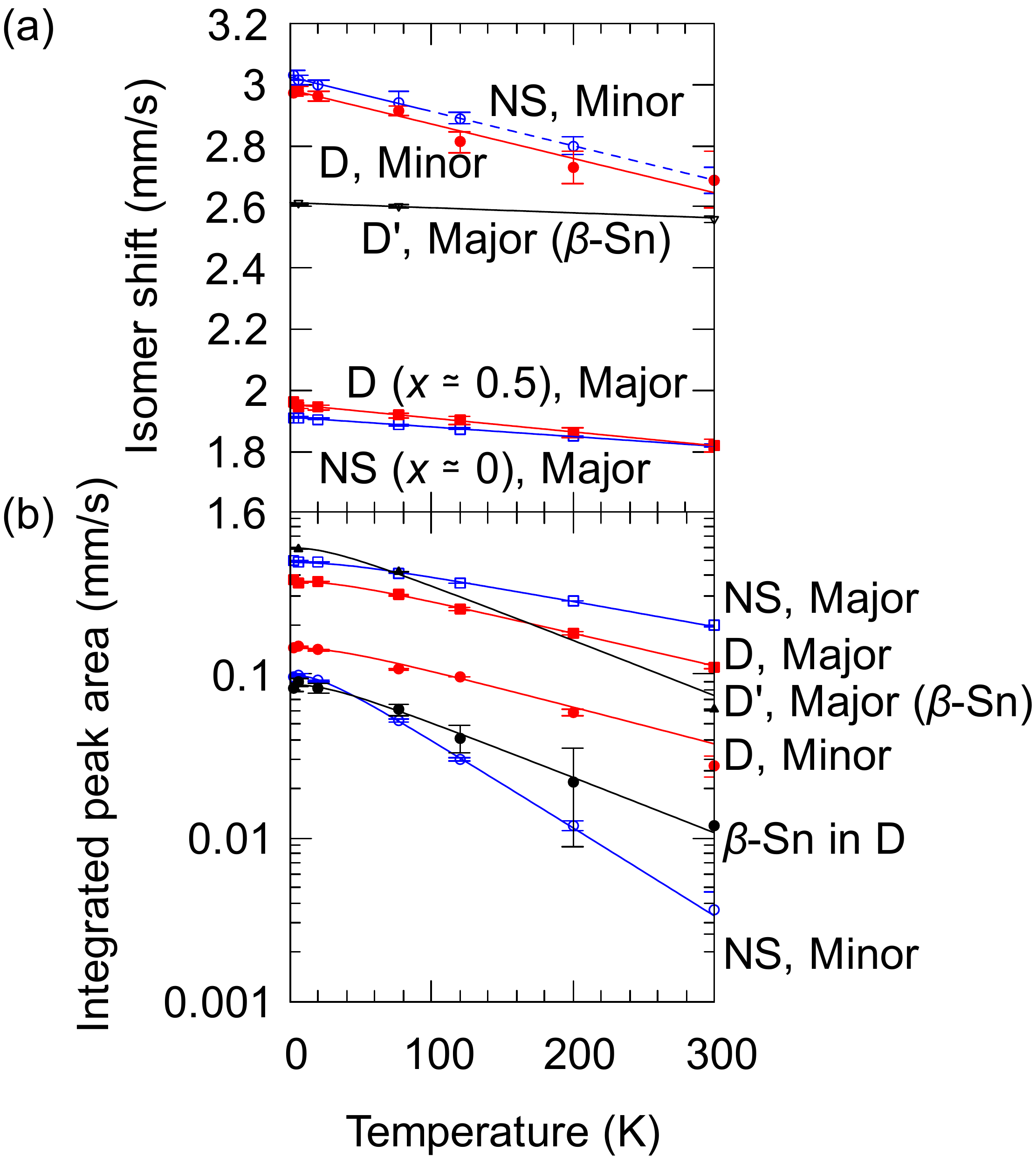}
\caption{(color online) Temperature dependence of (a) the isomer shifts and 
(b) integrated peak areas obtained by fitting the spectra with \revrev{multiple} Lorentzians.
\revrev{For the contribution from the $\beta$-Sn impurity in the D ($x\simeq0.5$) sample, we assumed 
that the integrated peak area exhibits the same temperature dependence as that of $\beta$-Sn contained in the D' sample.}
For the NS \rev{($x\simeq0$)} sample and at temperatures above 100~K, the double-Lorentzian fitting 
with six independent parameters (two sets of the peak position, intensity, and width) was not successful 
because of the too small contribution from the minor absorption.
\revrev{We estimated the isomer shift of the minor absorption by the following way and fixed the shift in the fitting 
(i.e. now we used five independent parameters).
First, we assumed that the isomer shift of the minor absorption exhibits a linear temperature dependence.
Then the slope of the temperature dependence is assumed to be the same as that of the D sample.
Under these assumptions, we extrapolated the data of the NS sample below 100~K to higher temperatures
to estimate the isomer shift at 120~K, 200~K, and 300~K, as indicated with the blue broken line in the panel (a).}}
\label{fig: isomer shift}
\end{figure}

\revrev{T}o examine the temperature evolution of the M\"{o}ssbauer effect, 
we plot the temperature dependence of the isomer shift and integrated intensity in Fig.~\ref{fig: isomer shift}.
The isomer shifts of the major absorptions 
are almost equal in the NS \rev{($x\simeq0$)} and D \rev{($x\simeq0.5$)} samples.
This fact suggests that the strontium deficiency affects so locally to the density of electrons at the nucleus position 
that a major fraction of the tin atoms remains unchanged from \ce{Sn^{4-}} even in the D sample.

\begin{table}
\caption{Isomer shifts (IS) at \revrev{300~K} and \revrev{effective} Debye temperatures $\left(\Theta\sub{D}\right)$ 
characterizing the vibration of the Sn atoms\revrev{,} obtained from experiments and calculations.
The accuracy of the experimental isomer shifts is expected to be about 0.01~mm/s.
The error of the experimental values are defined as the standard error of double-Lorentzian fitting.
As for the major absorption of the NS \rev{($x\simeq0$)} sample, the standard error of the fitting is smaller than the expected accuracy.
The accuracy of the calculated isomer shifts is around 0.2~mm/s.}
\label{tbl: IS and thetaD}
\begin{ruledtabular}
\begin{tabular}{ccc}
Absorption & IS (mm/s) & $\Theta\sub{D}$ (K) \\
\hline
Major \rev{[}NS \rev{($x\simeq0$)} sample\rev{]} & 1.82 & \revrev{222(3)} \\
Major \rev{[}D \rev{($x\simeq0.5$)} sample\rev{]} & 1.82(\revrev{2}) & \revrev{196(2)} \\
\SSO (calculation) & \revrev{1.7} & 186.5(8) \\
\hline
Minor (NS sample) & \revrev{2.69(4)\footnotemark} & \revrev{119.8(6)} \\
Minor (D sample) & \revrev{2.69(9)} & \revrev{185(7)} \\
\revrev{$\beta$-Sn in the D' sample} & 2.56(1) & 151(8) \\
``\StwoSO'' (calculation) & \revrev{3.1} & 97(3) \\
``\SthreeSO'' (calculation) & \revrev{2.8} & 148.5(18) \\
``\SfourSO'' (calculation) & \revrev{2.4} & 148.3(13)
\end{tabular}
\end{ruledtabular}
\raggedright
\footnotetext[1]{\revrev{Evaluated by linear extrapolation from the data at low temperatures (not by the Lorentzian fitting):
extrapolation error is presented.}}
\end{table}

\rev{From the temperature dependence of the spectra, we can calculate the energy of the local vibration of a Sn atom.
Since the thermal vibration of an atom changes the relative energy of the $\gamma$ ray by the Doppler effect, 
the intensity of the resonant absorption is suppressed at high temperature.
Therefore, the temperature dependence of the intensity reflects the thermal vibration and 
\revrev{provides} information on the binding \revrev{strength between} the atom \revrev{and} the lattice.}
The \revrev{observed} temperature dependence of the integrated intensity \revrev{is} plotted in Fig.~\ref{fig: isomer shift}(b)\revrev{.
Assuming that the intensity} is proportional to the recoil-free fraction $f$, 
\revrev{the data} was fitted with the Debye model~\cite{Bahgat1981Einstein}:
\begin{align}
f &= \exp\left(-\frac{\left(2\pi\right)^2\braket{\revrev{r}^2}\sub{D}}{\lambda^2}\right), \\
\braket{\revrev{r}^2}\sub{D} &= \frac{3\hbar^2}{Mk\sub{B}\Theta\sub{D}} \left[\frac{1}{4} + \left(\frac{T}{\Theta\sub{D}}\right)^2 
\int_0^{\left(\Theta\sub{D}/T\right)}\frac{y \mathrm{d}y}{\mathrm{e}^y-1}\right],
\end{align}
where $\braket{\revrev{r}^2}\sub{D}$ is the mean square displacement in the Debye solid, 
$\lambda=0.51933$~nm is the wavelength of the $\gamma$ ray, $\hbar$ is the reduced Planck constant, $M$ is the mass of a \ce{^{119}Sn atom}, 
$k\sub{B}$ is the Boltzmann constant, and $\Theta\sub{D}$ is the \revrev{effective} Debye temperature 
\revrev{(or M\"{o}ssbauer temperature)} characterizing the vibrations of the Sn atoms.
As shown in Fig.~\ref{fig: isomer shift}(b), the fitting reproduces well \revrev{the temperature dependences of} all absorptions.
The extracted \revrev{$\Theta\sub{D}$}, as well as the isomer shifts, 
are summarized in Table~\ref{tbl: IS and thetaD}.
Comparing \revrev{$\Theta\sub{D}$} of all absorptions, there is a tendency that \revrev{$\Theta\sub{D}$} decreases as the isomer shift increases.
This tendency can be \revrev{again} understood as the effect of the local strontium deficiency:
the higher local strontium deficiency corresponds to the lower density of the $p$ electrons at the tin atoms and weaker binding of the atoms to the lattice.

We also tried fitting with the Einstein model~\cite{Bahgat1981Einstein}:
\begin{equation}
\braket{\revrev{r}^2}\sub{E} = \frac{\hbar^2}{2Mk\sub{B}\Theta\sub{E}} \coth\left(\frac{\Theta\sub{E}}{2T}\right),
\end{equation}
where $\Theta\sub{E}$ is the \revrev{effective} Einstein temperature.
The fitting is as good as that with the Debye model but has slightly larger residual sum of squares.
We found that $\sqrt{3}\Theta\sub{E}$ gave almost the same temperature as $\Theta\sub{D}$.
This agrees with the theoretical relation $\sqrt{3}\Theta\sub{E}=\Theta\sub{D}$ deduced 
under the assumption of the same mean square displacements in the Einstein and Debye solids at high temperature: 
$\displaystyle{\braket{\revrev{r}^2}}\sub{E}=\displaystyle{\braket{\revrev{r}^2}}\sub{D}$~\cite{Bahgat1981Einstein}.

\revrev{Before closing this subsection, we would like to comment on the M\"{o}ssbauer spectrum in the superconducting state.}
We did not observe any clear difference in the spectra across the superconducting transition as shown in Fig.~\ref{fig: Moessbauer}(\revrev{e}).
Since superconductivity does not \revrev{cause large} change \revrev{in} the electronic density at the nucleus, 
the superconducting transition is in most cases difficult to be observed with M\"{o}ssbauer spectroscopy
in terms of the isomer shift~\cite{Bolz1975MoessbauerShift} and the recoil-free fraction~\cite{Hohenemser1965MoessbauerFraction}.
\revrev{To observe superconductivity in \SxSO in the M\"{o}ssbauer spectroscopy, we need to enhance the resolution as well as to use} 
samples with a larger superconducting volume fraction\revrev{. These are} the focus of future work\revrev{s}.

%


\subsection{Comparison with first-principles calculation}
\begin{figure}
\includegraphics[width=\linewidth]{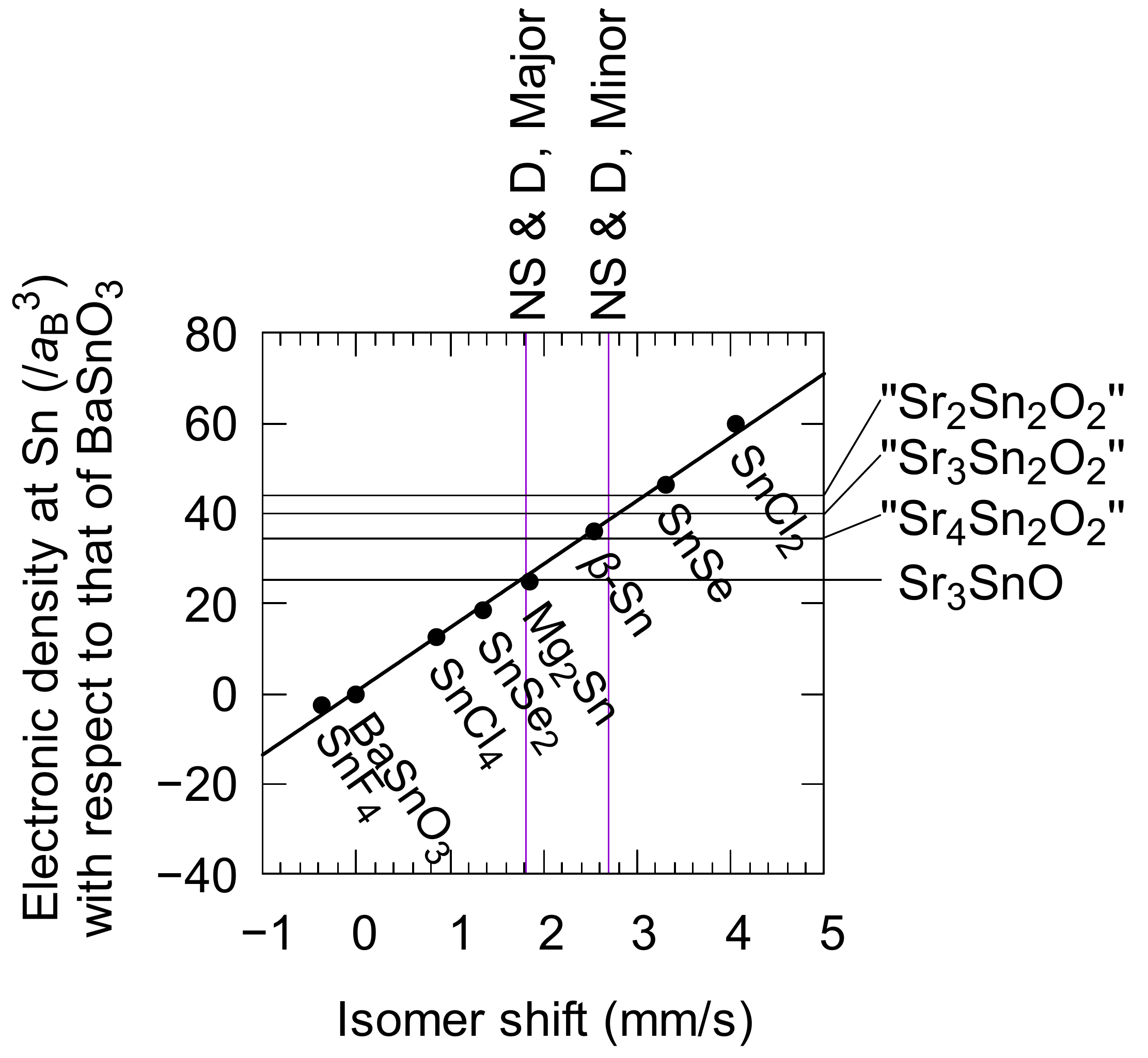}
\caption{(color online) Calculated electronic density versus experimental isomer shifts 
at room temperature~\cite{Stevens1983Mg2SnMoessbauer, Fournes1986SnF4Moessbauer}.
The thick solid line shows \revrev{the result of} the linear fitting \revrev{to} the data points.
The thin horizontal and vertical lines indicate the calculated electronic densities at the Sn nucleus of various \SSO-based \revrev{structures}
and observed isomer shifts of the NS \rev{($x\simeq0$)} and D \rev{($x\simeq0.5$)} samples, respectively.}
\label{fig: IS vs density}
\end{figure}

The relation between the experimental isomer shift (IS) at room temperature~\cite{Stevens1983Mg2SnMoessbauer, Fournes1986SnF4Moessbauer} 
and calculated electronic density at the Sn-nucleus position for various compounds are summarized in Fig.~\ref{fig: IS vs density}.
We find an anticipated linear relation~\cite{Fujita1999Moessbauer}:
$\textrm{IS} = \alpha\,\Delta\rho + \beta$ with $\alpha=0.071(2)a\sub{B}^3$~mm/s and $\beta=-0.03(6)$~mm/s, 
where $\Delta\rho$ is the electronic density with respect to that of \revrev{\ce{BaSnO3}} (262122.8~$a\sub{B}^{-3}$ in our calculation).
The coefficient $\alpha$ is smaller than the previously reported values by 10--20\%~\cite{Svane1997hyperfine, Kurian2009calibration}.
This discrepancy should originate from the difference in the calculation methods.

From this relation, the isomer shift can be estimated from the electron-density calculation with the accuracy of around 0.2~mm/s.
The calculated isomer shifts of \SxSO are listed in Table~\ref{tbl: IS and thetaD}.
The isomer shift of the major absorption agrees with the calculation for \SSO.
For the minor absorption, we calculated the density of electrons assuming various hypothetical arrangements of the Sr deficiency
and found that the calculated isomer shift of ``\SthreeSO'' [Fig.~\ref{fig: crystalline_structures}(c)] 
reasonably reproduces the observed values of the minor absorptions.

\begin{figure}
\centering
\includegraphics[width=\linewidth]{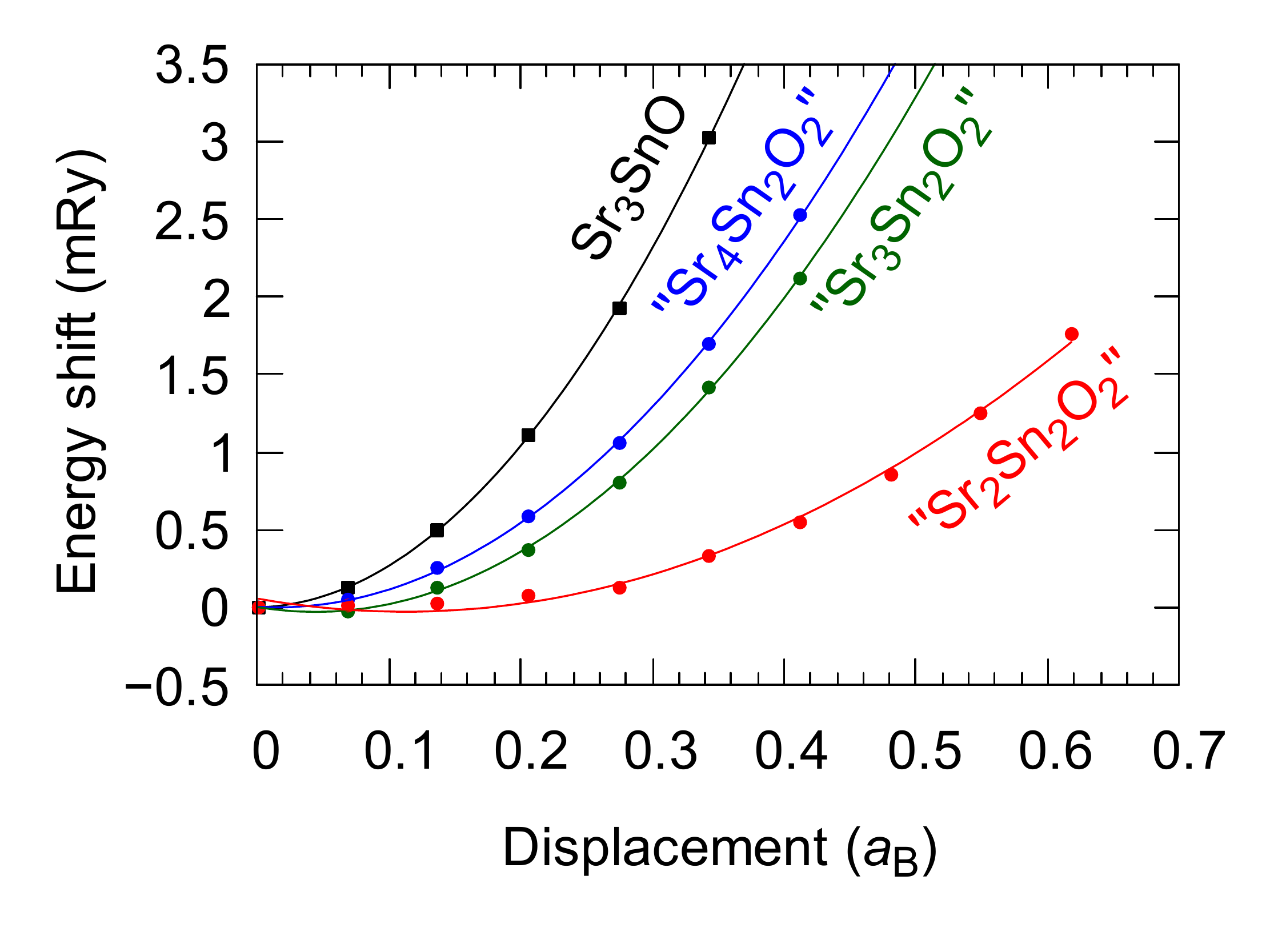}
\caption{(color online) Total \revrev{energies} of \SSO, ``\SfourSO,'' ``\SthreeSO,'' and ``\StwoSO'' 
as functions of the displacement of the Sn atom \revrev{calculated using the WIEN2k program}.
More Sr-deficient structures exhibit smaller changes in the total energy with displacement of the Sn atom,
meaning that these materials have lower force constants of the lattice.
In the highly deficient structures, the relative energy once goes negative with a small displacement\revrev{.
This means that the} original atomic position (zero displacement) \revrev{is} unstable.
Therefore, we fitted the data with the function $\Delta E = k(\revrev{r}-\revrev{r}_0)^2/2+E_0$ with the stable atomic position $\revrev{r}_0$, 
the force constant $k$, and the \revrev{minimum} energy $E_0$ as fitting parameters.}
\label{fig: pos vs ene}
\end{figure}

The force constant $k$ of the \revrev{Sn} atom in \SxSO \revrev{was} estimated from the change in the total energy due to a virtual displacement 
of the Sn atom \revrev{toward} the nearest neighboring strontium atom (See the arrows in Fig.~\ref{fig: crystalline_structures}):
$\Delta E = k(\revrev{r}-\revrev{r}_0)^2/2 + E_0$, where $\Delta E$ is the change in energy, $\revrev{r}$ is the displacement of the Sn atom, 
$\revrev{r}_0$ is the stable atomic position along the direction of displacement, and $E_0$ is the \revrev{minimum} energy corresponding to $\revrev{r}=\revrev{r}_0$.
The effective Debye temperature was then calculated by 
$\Theta\sub{D} = \sqrt{3}\Theta\sub{E} = \sqrt{3}\left(\hbar/k\sub{B}\right)\sqrt{k/M}$~\cite{Bahgat1981Einstein}.
Figure~\ref{fig: pos vs ene} shows the change in energy and \revrev{results of the fitting. 
The} extracted \revrev{effective} Debye temperatures are listed in Table~\ref{tbl: IS and thetaD}.
\revrev{The overall trend of the decreasing $\Theta\sub{D}$ with the increas\revrev{e of the} isomer shift is reproduced in calculation.}
\revrev{The} calculated \revrev{$\Theta\sub{D}$} of \SSO \revrev{and} the experimental \revrev{$\Theta\sub{D}$ 
of the major absorption of the NS ($x\simeq0$) sample differs by 16\%.
For the minor absorptions, the calculated $\Theta\sub{D}$} of ``\StwoSO'' \revrev{and ``\SthreeSO'' 
matches by similar accuracy with the experimental $\Theta\sub{D}$} 
of the minor absorption\revrev{s} in the NS \revrev{and D ($x\simeq0.5$)} sample\revrev{s, respectively}.

These good agreements between the calculated and experimental values confirms 
that the origins of the major and minor absorptions are the tin atoms without and with the neighboring strontium deficiency, respectively.
We note that the local deficiency is more in the NS \rev{($x\simeq0$)} sample 
than in the D \rev{($x\simeq0.5$)} sample \revrev{based on the estimation of $\Theta\sub{D}$ of the minor absorptions}.
This may be due to the slow cooling in the end of the synthesis of the NS sample;
the stoichiometric \SSO solidifies first and \SxSO with a large deficiency is formed later as a minority phase.
In the case of the D sample, in contrast, \revrev{relatively rapid cooling} leads to a large amount of deficient \SxSO but with relatively dilute deficiency.

\begin{figure}
\includegraphics[width=0.8\linewidth]{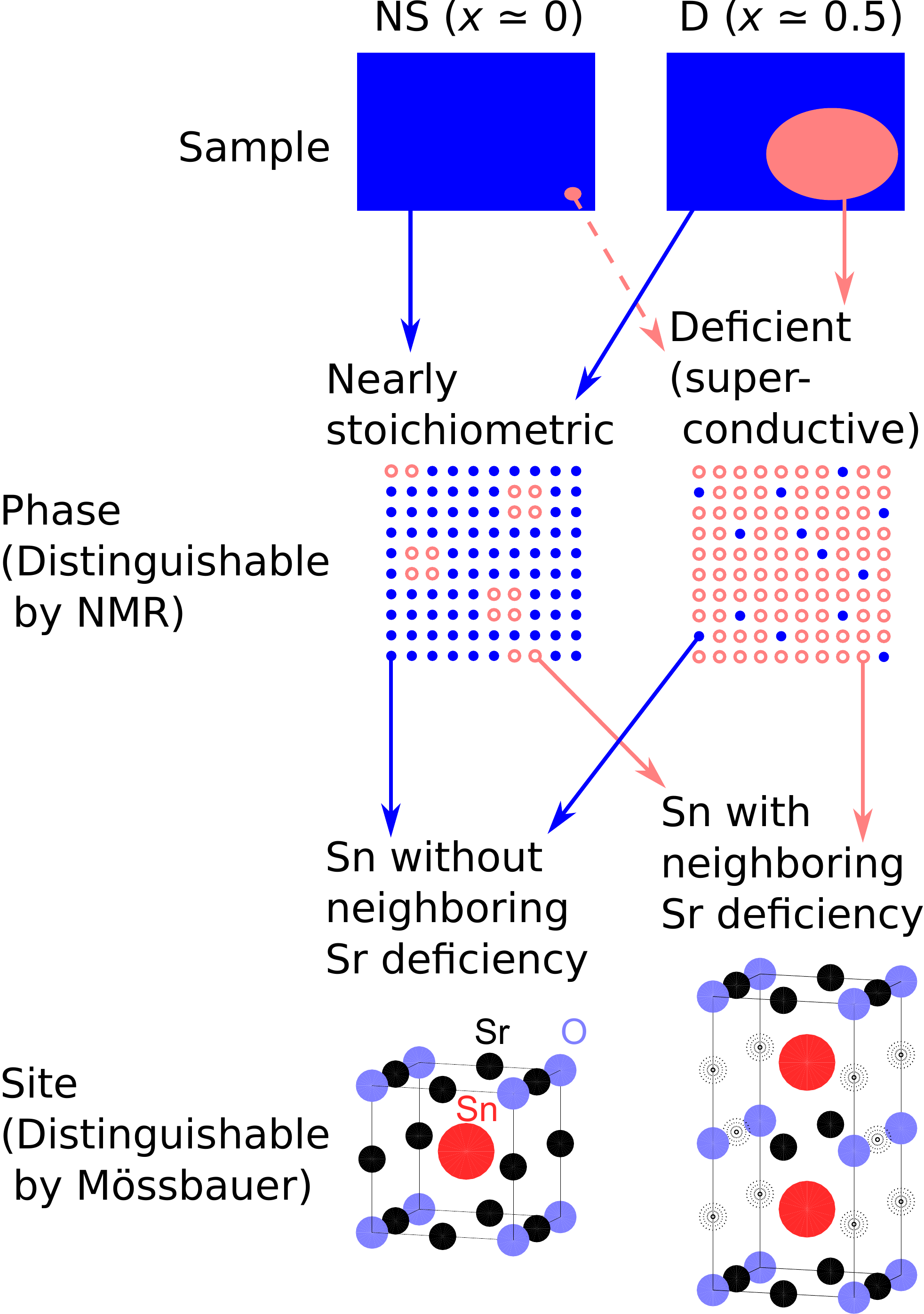}
\caption{\revrev{(color online) Schematic illustration of the phase and site splittings.
The NS sample is dominated by the nearly stoichiometric phase (blue) 
possibly with a separated inclusion of the deficient phase (red).
In contrast, the D sample contains about 10\% of the deficient superconductive phase.
The nearly stoichiometric phase consists mostly of the Sn atom without the neighboring deficiency of Sr (blue solid circles)
while the superconductive phase is composed mainly of the Sn atom with neighboring deficiency (red open circles).}}
\label{fig: splittings}
\end{figure}

\subsection{Comparison with NMR}
We comment that the phase splitting into the nearly stoichiometric and deficient compounds has been 
suggested in the previous study of magnetization~\cite{Oudah2019SrDeficiency} and indeed observed with NMR~\cite{Kitagawa2018Sr3-xSnO}.
Since NMR probes \revrev{properties of} the conduction electrons though it measures the Zeeman energy of the Sn nucleus, 
the splitting in the NMR spectrum suggests that one sample consists of two materials \revrev{(``phases'')} 
with distinct stoichiometries \revrev{as shown in Fig.~\ref{fig: splittings}}.
\rev{Similar phase splitting must exist also in the current samples, judging from the shielding fraction lower than 100\%.}
On the other hand, the M\"{o}ssbauer spectroscopy detects the local information at the nucleus position, 
and thus the two absorptions may originate from two sites in \rev{each phase}.
Since highly deficient phases such as ``\SthreeSO'' and ``\StwoSO'' will be chemically unstable,
it is unlikely that one sample separates into the two phases \SSO and ``\SthreeSO.''
Thus, we conclude that the minor absorption reflects the local Sr deficiency in \rev{each of the phases contained in} 
\SxSO with $x\simeq0$ in the NS sample and with $x\simeq0.5$ in the D sample.

\rev{To sum up, we consider that our samples contain nearly stoichiometric and deficient phases \revrev{as depicted in Fig.~\ref{fig: splittings}}.
Each of these phases contain\revrev{s} Sn sites with and without neighboring Sr deficiencies.
Previous NMR measurements observed the phase splitting but could not detect such site splitting within each phase.
In contrast, the M\"{o}ssbauer spectroscopy detects the difference in the Sn sites with and without neighboring Sr deficiency 
but hardly distinguishes similar sites in different phases.}

\rev{The present results confirm the \ce{Sn^{4-}} state and thus the validity of the band structure calculations.
In this sense, these results are consistent with the NMR experiments observing the $1/T_1$ attributable to the Dirac dispersion~\cite{Kitagawa2018Sr3-xSnO}.
Furthermore, since the large amount of the Sr deficiency to ``\SthreeSO'' or ``\StwoSO'' leads to heavy hole doping,
the deficient phase with a large portion of such Sn sites with neighboring Sr deficiency would exhibit metallic $1/T_1$, as indeed observed in
the NMR measurements.}

\section{Conclusion\label{conclusion}}
By combining the M\"{o}ssbauer spectroscopy and first-principles calculations, we confirmed the \ce{Sn^{4-}} state in the antiperovskite oxide \SxSO.
Furthermore, we identified the origin of the minor absorption to be the Sn atoms with the neighboring Sr deficiency.
From the effective energy of the local lattice vibrations, we \revrev{infer} that the local Sr deficiency 
is more in the non-superconductive sample\revrev{, implying} that the moderate deficiency 
between ``\SfourSO'' and ``\SthreeSO'' is needed for superconductivity in \SxSO.
\rev{As a future experiment, measurements of the Sr-Sr distance using a pair-distribution function 
will further clarify the local environment of the Sr deficiency.}
We hope that the microscopic understanding of the \revrev{novel ionic} states will lead to new chemistry and physics of metallic anions
and will also contribute to development of the M\"{o}ssbauer spectroscopy.


\begin{acknowledgments}
We acknowledge J. N. Hausmann and I. Markovi\'{c} for their contribution in preparation of the samples.
We thank Y. Kobayashi and R. Masuda for the support and discussion regarding the M\"{o}ssbauer spectroscopy.
We are grateful to M. Kawaguchi for his contribution in revision of the manuscript.
This work was partially supported by Research Center for Low Temperature and Materials Sciences in Kyoto University, 
by Japan Society for the Promotion of Science (JSPS) KAKENHI Nos.\ JP15H05851, JP15H05852, JP15K21717 (Topological Materials Science), 
JP17H04848, and JP17J07577, and by the JSPS Core-to-Core Program (A. Advanced Research Network), 
as well as by Izumi Science and Technology Foundation (Grant No.\ H28-J-146).
AI is supported as JSPS Research Fellow.
\end{acknowledgments}

%

\end{document}